# Hybrid Longitudinal–Transverse Propagating Electric Fields in Photonic Crystal Waveguides

**Yanrong Zhang[1], Hooman Barati Sedeh[2], Christopher S. Whittington[3], Natalia M. Litchinitser[2], Shuren Hu[1*], Sharon M. Weiss[1*]**

*1.Department of Electrical and Computer Engineering, Vanderbilt University, Nashville, Tennessee 37235, USA*

*2.Department of Electrical and Computer Engineering, Duke University, Durham, North Carolina 27708, USA*

*3.Interdisciplinary Materials Science Program, Vanderbilt University, Nashville, TN 37235, USA*

** Contact author e-mail addresses: shuren.hu@vanderbilt.edu, sharon.weiss@vanderbilt.edu*

**Abstract**

In a uniform, source-free, and unbounded medium, Maxwell's equations require electromagnetic waves to be purely transverse. However, when a beam of light is tightly focused or strongly confined, a longitudinal field component can emerge[1–3]. Strong longitudinal fields enable many novel phenomena and applications, including single molecule-detection[4,5], near-field imaging[6,7], and high-resolution photolithography[8]. Although the behavior of the longitudinal electric (LE) field component of the electromagnetic field in ordinary waveguides is well established, judicious nanostructuring offers unprecedented control over its strength as well as spatial and spectral distribution. Here, we demonstrate a full-vectorial theory and experimental results showing that for specially designed waveguides, such as one-dimensional antislot photonic crystal (PhC) waveguides, the LE field can hybridize with the transverse electric (TE) field in the waveguide and can be subsequently decomposed into independent polarizations through far field imaging. When the in-plane mirror symmetry of a PhC unit cell is broken, coupling between LE and TE modes produces two hybrid LE-TE modes and opens a new photonic bandgap. The LE–TE composition of the hybrid modes and the width of the resulting bandgap can be tuned by changing the rotation angle of the antislot within the unit cell. We show that a 45° antislot orientation with respect to the propagation direction yields hybrid modes with the largest LE field contribution and the widest geometry-induced bandgap. Such engineered PhC waveguides enable new on-chip photonic functionalities, including in-plane angle-invariant dipole coupling in quantum systems, higher-order polarization-division multiplexing, and enhanced control of light flow.

## Introduction

A purely transverse description of electromagnetic waves is incomplete when light is strongly confined or the medium is non-uniform and bounded, such as in tightly focused beams and guided-mode waveguides. In these regimes, the longitudinal electric field must also be taken into account. Longitudinal field components were first noted by Maxwell[9] and were later analyzed explicitly by Rayleigh[10], Heaviside[11], and Sommerfeld[12]. The contribution of LE field components was investigated after scalar approximations were found to be inadequate for high numerical-aperture focusing[13]. Instead, a full-vectorial model that considers the rotation of polarization caused by a high-numerical aperture was required to accurately represent the behavior of light in this regime[14,15]. This vectorial description facilitated both theoretical and experimental studies of the LE field component, which demonstrated that the LE field produces an on-axis intensity distribution at the focal point[1]. Building on this insight, subsequent work has focused on engineering and enhancing the longitudinal field component through the use of annular apertures[8], high numerical aperture focusing[16] and radially polarized beams[4,17], enabling novel applications such as near-field probes[18] and nonlinear frequency conversion[19] that otherwise would not have been possible.

For guided-wave systems, similarly, a semi-vectorial approximation fails to capture the complex relationships between $E_x$, $E_y$ and $E_z$ field components for a confined electromagnetic field. Consequently, guided-wave phenomena are typically analyzed using full-vectorial finite-difference–based mode solvers, which accurately evaluate field distributions and modal polarization mixing in dielectric waveguides[20]. In particular, numerical studies of silicon nanowire waveguides revealed that the LE field component can reach a relative amplitude nearly equivalent to that of the dominant TE component[3]. Other studies have applied the full-vectorial framework to optical microfibers[21] and nanopillars[22], where the strong confinement produces substantial LE fields that influence coupling, propagation loss, second-harmonic generation and spin-orbit interactions. In traditional waveguides, the strength of the LE field component can be controlled but the spatial and spectral distribution of the LE field cannot be easily tuned; the LE field typically has maximum amplitude at the sidewalls[23,24] , surfaces, or terminated end of the waveguide and is out-of-phase with respect to the magnetic field[24]. Thus, the LE field component within these systems only oscillates locally, which does not contribute to the net energy transport[3]. However, according to Gauss's law, introducing a spatially varying permittivity along the propagation direction allows LE fields to be partially in-phase with the magnetic field, which allows the LE field to propagate together with the TE field. Therefore, although LE fields in guided structures, high numerical aperture focusing, and nonuniform media are well established and extensively studied, their controlled manipulation remains an open challenge. Judicious nanostructuring of waveguides at subwavelength scale along the propagation direction provides a pathway to overcome this challenge.

In this work, we demonstrate through theory, numerical simulations, and experiments that a subwavelength-engineered PhC waveguide, comprising antislot unit cells[25] with appropriately designed refractive index modulation, supports both LE and TE dominated modes that can be hybridized by intentionally breaking the mirror symmetry of the unit cell by rotation (Figure 1). The coupling strength between the LE and TE modes, as well as their spatial and spectral distribution, can be tuned by changing the angle of the antislot unit cell. We demonstrate that nanostructuring of the waveguide relocates the LE field to the waveguide center and enables deterministic control of the polarization of the hybrid LE–TE mode. We believe this work can open new avenues for both fundamental research and practical applications. The in-plane polarization insensitivity of hybrid LE-TE modes provides a potential mechanism for on-chip angle-invariant coupling of 2D material emitters. Moreover, the ability to tune the coupling strength between the LE and TE modes could be exploited for high-density data transmission, enabling a novel form of polarization division multiplexing that promises to significantly increase the bandwidth of on-chip optical interconnects.

**Origin of the LE mode**

For a waveguide that is homogeneous along the propagation direction ($x$-direction), such as a traditional ridge waveguide, the transverse fields are in-phase with respect to each other and give rise to energy transport along the $x$-direction. However, according to Gauss's law, as shown in Supplementary Note 1, the LE field ($E_x$) is out-of-phase with the TE field ($E_t$) and does not contribute to the time-averaged power flow, instead storing reactive electromagnetic energy. The magnitude of the LE field scales with the spatial gradient of $E_t$ and is generally not considered except for special cases such as nanowire waveguides that support strong transverse field gradients[3].

Re-examining Gauss's law under the condition that the permittivity of a waveguide medium can vary along the propagation direction reveals a more general relationship between $E_x$ and $E_t$, as shown in Eqn. 1 and derived in Supplementary Note 1. For a homogeneous waveguide, $\frac{\partial \varepsilon}{\partial x} = 0$ and $E_x$ is purely out-of-phase with $E_t$, while for a waveguide with a non-homogeneous dielectric constant along the direction of propagation, such as a PhC waveguide, $\frac{\partial \varepsilon}{\partial x} \neq 0$ and $E_x$ is partially in phase with $E_t$. In this perspective, since the LE field is not completely out-of-phase with the TE field components, the in-phase LE field can contribute to the propagating mode as

$$E_x \propto \frac{\partial \varepsilon}{\partial x}\left(\nabla_t \cdot (\varepsilon E_t)\right) + i\varepsilon\beta\nabla_t \cdot \varepsilon E_t \quad (1)$$

While Gauss's law specifies the relationships between the electric field components, Ampère's law uniquely relates the $E_x$ and $E_y$ fields to the $H_z$ field. As shown in Supplementary Note 2 and Fig. S1, for a given complex $H_z$ distribution, the transverse and longitudinal components of the electric field can be given by $E_y = \left(-\frac{1}{i\omega\varepsilon}\right)\left(\frac{\partial H_z}{\partial x}\right)$ and $E_x = \left(\frac{1}{i\omega\varepsilon}\right)\left(\frac{\partial H_z}{\partial y}\right)$, respectively. Because $H_z$ is the unifying field component when considering both $E_x$ and $E_y$, we analyze our PhC waveguides using band structure calculations (Fig. 1) with an $H_z$ excitation source in MEEP[26]. This $H_z$ excitation produces bands from both TE and LE modes. To understand the origin of each band, the band structure calculation was performed with a Gaussian beam source in two different polarization configurations: (i) an electric field along the $y$ direction and (ii) an electric field along the $x$ direction. While the $H_z$ source excites every supported mode or band, the two separate sources each only excite bands according to their supported polarizations. The bands strongly excited from an $E_y$ polarization are identified as TE-dominant bands, while the band that is strongly excited by the $E_x$ polarization is identified as the novel LE-dominant band. We note that while it is straightforward to introduce an $E_x$ polarized source in simulation, it is not so straightforward to introduce such an excitation light source in experiments as typical excitation sources produce light with the electric field oriented transverse to the propagation direction. We also note that in traditional one-dimensional PhC studies[27], TE modes are classified into branches based on the symmetry of the field distribution with respect to the y = 0 reflection, reporting band structures similar to those shown in Fig. 1a. Typically, only the lower two TE modes are examined, where the magnetic field exhibits even symmetry and produces a field profile similar to the left panel of Fig. 1a.

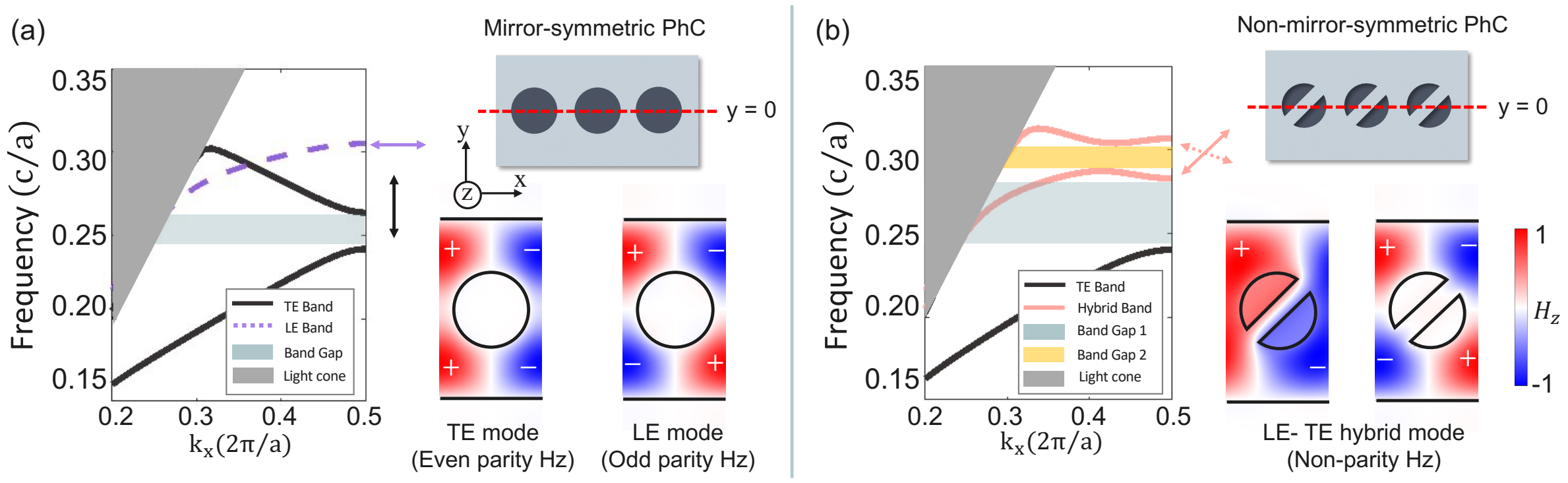

Figure 1. TE and LE modes and their hybridization in PhC waveguides. (a) Schematic illustration and band structure of 1D PhC waveguide with mirror-symmetry in the x-y plane. Solid black lines in the band diagram are TE bands, which are strongly excited by an electric field polarized in the y-direction or a magnetic field polarized in the z-direction, while the dashed purple line is an LE band, which is strongly excited by an electric field polarized in the x-direction or a magnetic field polarized in the z-direction. The magnetic field ($\vec{H}$) profiles of the TE and LE modes at the band edge are also shown. (b) Schematic illustration and band structure of 1D PhC waveguide without mirror-symmetry across the y=0 plane. Two hybrid bands (solid red lines) are excited by any in-plane electric field polarization or magnetic field polarized in the z-direction. The corresponding magnetic field profiles of the band-edge modes are shown, revealing two distinct nonsymmetric LE -TE hybrid modes.

In our work, we further identify a higher-order TE mode whose magnetic field exhibits odd parity under y = 0 reflection in the x-y plane, which is accompanied by a strong LE field component. Because of this notable field contribution, we therefore refer to this mode as the LE mode. Moreover, in contrast to previously reported photonic band structures, we judiciously engineer the LE band to shift downward and intersect with the conventional TE band by modifying the hole size in the PhC unit cell and waveguide width, thereby creating a crossing point in k-space, which enables coupling of these modes when the mirror symmetry across y = 0 in the PhC unit cell is broken (Fig. 1b), as described in the next section.

**Hybridization between LE and TE modes**

In conventional waveguides, symmetry breaking in the y-z cross-section couples TE and TM modes by lifting their symmetry-based orthogonality[28]. An analogous mechanism operates in the x-y propagation plane of a PhC waveguide, where symmetry breaking enables coupling between LE and TE modes. Such symmetry breaking in the x-y plane can be achieved through PhC unit-cell design that breaks the mirror symmetry responsible for the odd- and even-parity magnetic-field distributions of the LE and TE modes. Here, we implement this approach using a silicon antislot PhC waveguide, whose unit cells consist of circular air holes bridged by a silicon bar (i.e., antislot; Fig. 1b, Fig. 2a,b)[25]. Rotating the silicon antislot bar away from the x- or y-directions breaks symmetry in the x–y plane and induces LE–TE coupling, producing two hybrid modes with no definite $H_z$ symmetry (Fig. 1b). This coupling process is well-described using coupled mode theory (see Supplementary Note 3, Supplementary Tables 1-2 and Figs. S2-4) and produces an anti-crossing in the band structure that manifests as a new photonic bandgap. The rotation angle of the antislot directly controls the size of the resulting bandgap, as shown in Fig. 2c(ii–iv), which in turn indicates the strength of the LE–TE coupling. The coupled modes, referred to as the upper and lower hybrid band modes, exhibit a unique property that arises from their coupled nature and non-symmetric field distribution, such that both hybrid modes can be excited using light with any polarization. Importantly, this allows experimental measurement of the newly emerged photonic bandgap and hybrid LE-TE modes using excitation from a conventional laser source (analogous to an $E_y$ polarized source).

For the symmetric antislot PhC waveguides (i.e., R0 and R90 antislot PhC waveguides, where the notation R$\theta$ denotes the rotation angle $\theta$ in degrees of the antislot bar with respect to the propagation direction), TE and LE modes have negligible interactions between each dominant field, as shown in Fig. 2d, e and h, i. Through subwavelength engineering and consideration of interface conditions for electric fields, we can control the LE field distribution into two branches: LE field uniform within the center of the waveguide (i.e., antislot effect, where the tangential component of the electric field, $E_x$, is continuous) or LE field highly localized to the air region (i.e., slot effect, where the normal component of the electric displacement field is continuous across the interface and the electric field is enhanced in the lower dielectric slot region). In contrast, for antislot PhC waveguides with an intermediate antislot rotation angle for which the mirror symmetry across y = 0 is broken (e.g., R45), hybridization occurs and two hybrid TE–LE bands appear. At the band edges, where mode coupling is maximal (see Supplementary Fig. S2-3), both upper and lower hybrid modes exhibit significant $E_x$ and $E_y$ components, as shown in Fig. 2f,g. For example, the ratio of $|E_x|/|E_y|$ for the R45 antislot PhC upper and lower hybrid mode integrated over the entire unit cell is approximately 90%, indicating a strong mixed polarization character. In contrast, there is a clear distinction between the dominant electric field components of the TE and LE modes for the R0 and R90 antislot PhC waveguides, as shown in Fig. 2d,e and Fig. 2h,i, respectively. Supplementary Table 3 highlights the key differences between LE field contributions in various free space and guided wave configurations.

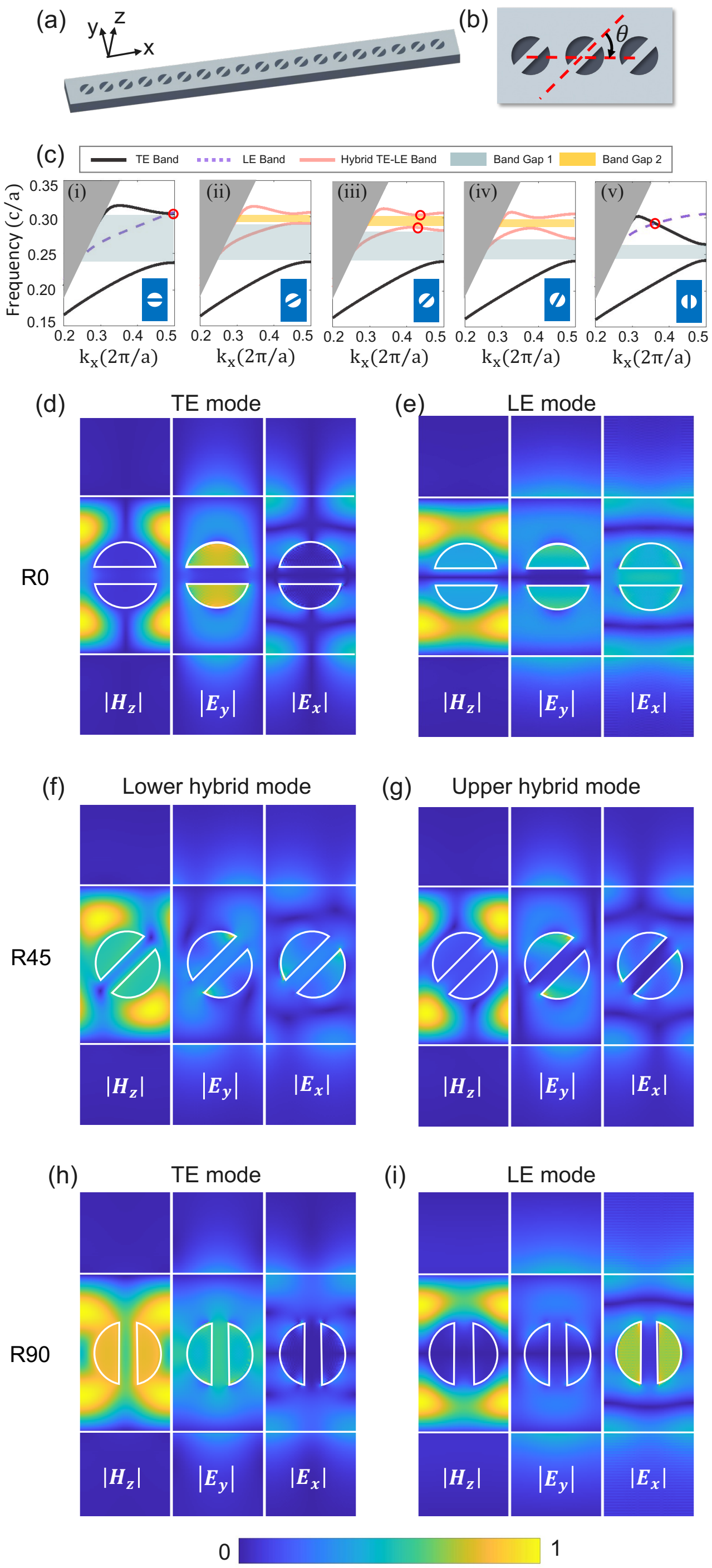

Figure 2. Broken mirror symmetry-enabled LE-TE hybridization and band evolution in silicon antislot PhC waveguides. (a) Schematic of R45 antislot PhC waveguide and (b) zoom-in on three unit cells. (c, i-v) Band diagrams

for R0, R30, R45, R60 and R90 antislot PhC waveguides. Black solid lines are TE bands, purple dashed lines are LE bands, and red solid lines are hybrid LE-TE bands. The blue shaded region represents the fundamental photonic band, which is caused by breaking the translational symmetry of the waveguide. The yellow shaded region is the novel photonic band gap, which is generated by breaking the *x-y* mirror symmetry in the unit cell. (d-i) For each panel, the plots represent 2D slices of $|H_z(x,y)|$, $|E_y(x,y)|$ and $|E_x(x,y)|$ field distributions within a single unit cell. The amplitude of $E_x$ and $E_y$ are normalized to the total electric field $E_{total}$. The labeled modes correspond to the circled points in the band structures. For R0 and R90 unit cells, the circled modes lie at the band-crossing point, while for the R45 unit cell, the circled modes correspond to the band edge.

## New photonic bandgap from hybrid LE-TE modes

To investigate the novel bandgap induced by hybridization between the LE and TE modes, we conducted finite-difference time-domain (FDTD) simulations and experimental transmission measurements on silicon antislot PhC waveguides. Fig. 3a shows a scanning electron microscope (SEM) image of an R45 antislot PhC waveguide (fabrication details provided in Methods). Fig. 3b (i) shows the simulated transmission spectra of antislot PhC waveguides with antislot rotation angles between 0° and 90°, presented as a heat map. At 0°, the bandgap is closed. As the antislot rotation angle increases, the bandgap progressively opens and reaches its maximum at 45° before narrowing as the antislot angle continues to increase to 90° when the bandgap is closed. This trend is consistent with the band diagrams of the LE and TE modes in Fig. 2. Moreover, experimental measurement of the transmission spectra of the antislot PhC waveguides with antislot rotation angles between 0° and 90° (Fig. 3b(ii)) also shows good agreement with the FDTD simulations. Fig. 3c shows a direct comparison of selected simulated and measured transmission spectra for R90, R75, R60, and R45 antislot PhC waveguides, clearly illustrating the evolution of the new bandgap formed from the hybridization of the LE and TE modes, and the close agreement between simulation and experimental results. These results support the coupled mode theory calculations (Fig. S2-S4) showing that the antislot rotation angle can be tuned to directly control the degree of coupling between the LE and TE modes: 0° and 90° correspond to the uncoupled limits while 45° represents the condition of maximum mode hybridization.

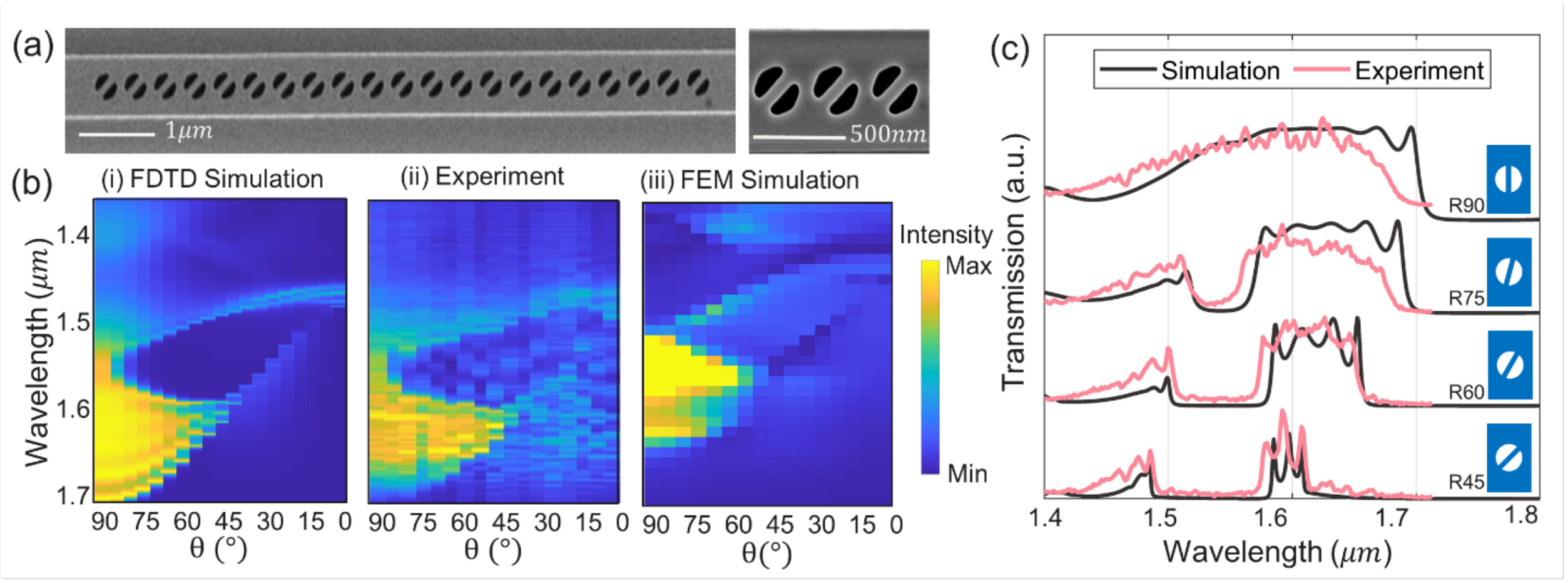

Figure 3. (a) SEM image of R45 antislot PhC waveguide and zoom in image of three unit cells. (b, i) Simulated and (b, ii) experimentally measured transmission spectra of antislot PhCs with different antislot rotation angles (R0 to R90) displayed as heat maps. (b,iii) finite element method (FEM) simulations of forward to backward scattered power based on the multipole decomposition approach. (c) Simulated (black) and experimentally measured (red) transmission spectra of R90, R75, R60 and R45 antislot PhC waveguides. The new broken mirror symmetry-enabled photonic bandgap is centered near 1.55 μm. As the PhC unit cell rotates away from the symmetric R0 and R90 configurations, the band gap opens and reaches its maximum bandwidth near R45, in good agreement between simulation and experiment.

Building on the rigorous PhC analysis above, we show that the fundamental light–matter interactions in antislot PhC waveguides can also be understood intuitively by viewing the structure as a chain of coupled resonant meta-atoms, with each meta-atom corresponding to a single antislot unit cell. For nonmagnetic meta-atoms, the optical response is governed by the induced electric polarization $P$ in arising from the applied electromagnetic field. Upon the interaction of light with a meta-atom, the induced polarization is related to the field distributions within the particle via $P_{in} = \varepsilon_0 \left(\varepsilon_p - \varepsilon_d\right) E_p$ , where $\varepsilon_0$, $\varepsilon_p$ and $\varepsilon_d$ are the dielectric constants of free space, the meta-atom, and the surrounding medium, respectively, and $E_p$ is the total electric field inside the meta-atom[29–31]. Therefore, the internal field distribution can directly change the induced polarization, which in turn alters the excited moments (i.e., dipole, quadrupole, etc.) within the scatterer (details in Supplementary Note 4). Using the multipole decomposition approach, we studied the ratio between forward and backward scattered power of a chain of seven meta-atoms as a function of wavelength and the antislot rotation angle in finite element method simulation, as shown in Fig. 2b (iii). The ratio of forward-to-backward scattered light closely follows the trends observed in both the band structure calculations and transmission spectra of the antislot PhC waveguides. The slight difference in spectral position of the bandgap is attributed to including only seven silicon meta-atoms instead of a larger array and neglecting the silicon dioxide substrate due to computational limitations. The directional response of the scattered power provides additional evidence of the LE–TE hybridization and its influence on wave propagation.

**Far-field scattering: LE and TE field components**

While the experimental observation of a new bandgap confirms mode hybridization, we are still interested in understanding the composition of the electric-field polarization. Therefore, to experimentally verify and quantify the LE field component relative to the TE field component, we measured the far-field scattering from silicon antislot PhC waveguides with rotation angles between 0° and 90° and compared to far-field scattering simulations (details provided in Methods). The results of these simulations and measurements are shown in Fig. 4. For the R0 and R90 antislot PhC waveguides, the far-field scattering profile is dominated by the $\left|E_y\right|^2$ (TE) component, which is consistent with the above analysis. As the rotation angle approaches 45°, the LE ($E_x$) component becomes increasingly significant. For R45 antislot PhC waveguides, the ratio of $|E_x|^2/|E_y|^2$ integrated over the far-field profile reaches approximately 50% in both simulation and experiment (see Extended Data Fig. 1 and Method). We anticipate that this ratio can be further increased using unit-cell geometries that support stronger electric and magnetic field gradients.

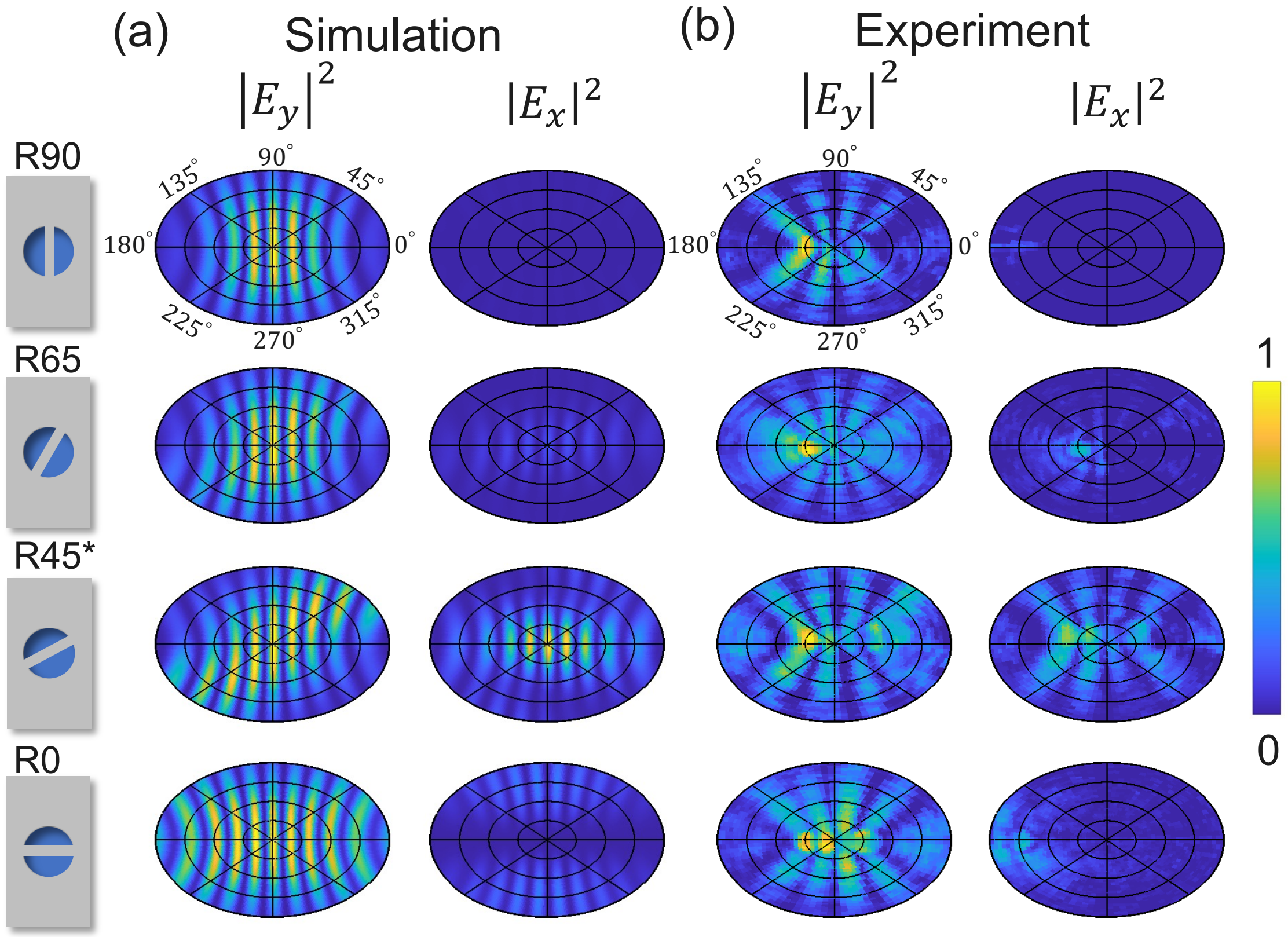

Figure. 4. Far-field scattering reveals the LE field contribution in rotated antislot PhC waveguides. (a) Far-field projection obtained from FDTD simulation and (b) out-of-plane far-field scattering measured experimentally from silicon antislot PhC waveguides with rotational angles of R90, R65, R45, and R0. For the R90 and R0 structures, the far-field response is dominated by the TE component, consistent with a TE-like guided mode. As the rotation angle approaches 45 degrees, the LE field component becomes increasingly significant, leading to a pronounced modification of the far field scattering profile. Note that the simulated far-field scattering of the R45 antislot PhC waveguide is compared against the measured far-field spectra from an R44 antislot PhC waveguide device. The far-field scattering observed near R45 confirms that the LE field possesses a non-negligible in-phase component and can be decomposed from the hybrid LE-TE mode, enabling its use as an independent in-plane polarization channel. All far-field intensities are normalized to the maximum value at each rotation angle of the antislot PhC waveguide.

Most importantly, the observed far-field scattering provides direct evidence that the LE component contains a non-negligible in-phase contribution. Because far-field radiation depends on both the amplitude and relative phase of the near-field components, a purely out-of-phase LE field would contribute only reactive energy and therefore radiate weakly. Instead, the pronounced far-field scattering observed for the R45 antislot PhC waveguide indicates that the LE field acquires a partial in-phase relationship with the magnetic field. Finally, our measurements show that the LE field component can be selectively decomposed from the hybrid TE–LE mode, enabling its use as an independent polarization channel. This capability opens a pathway to on-chip data communication schemes that exploit multiple in-plane polarization degrees of freedom.

**Conclusion**

In summary, we have demonstrated both theoretically and experimentally that a LE mode in a one-dimensional antislot PhC waveguide can be made directly accessible and measurable through broken mirror symmetry-enabled hybridization with a TE mode. By breaking the in-plane mirror symmetry of the PhC unit cell, two guided hybrid LE-TE modes are formed in which the longitudinal and transverse electric field components jointly contribute to the energy flow. Moreover, these hybrid modes open a novel bandgap through their mode coupling, whose width can be continuously tuned by the antislot rotation angle. At a rotation angle of 45°, the bandgap attains its maximum width and the coupling between the LE and TE modes is strongest. Evidence for hybrid LE-TE modes was shown through band structure calculations, simulated field profiles and transmission spectra, multipole decomposition scattering simulations, and transmission and far-field scattering measurements. The hybrid LE-TE modes could enable applications ranging from new dimensions of polarization multiplexing to higher coupling efficiency of quantum emitters embedded in waveguides. More broadly, this work establishes LE engineering as an additional degree of freedom in integrated photonics, opening avenues for novel device concepts in compact planar platforms.

## Methods

Antislot PhC waveguide fabrication

A series of antislot PhC waveguides with rotational angles varying from 0° to 90° were fabricated on a silicon-on-insulator wafer featuring a 270 nm silicon device layer and a 3 μm buried oxide layer. Standard electron beam lithography and reactive ion etching techniques were used to create these waveguides on a single chip within a single batch. The fabricated radius of the air holes of the PhC was 152 nm, the width of the antislot bar was 85 nm, the periodicity of the unit cell was 420 nm, the width and the thickness of silicon nanobeam waveguide were 740 nm and 270 nm, respectively, on top of the silicon dioxide layer, and the number of unit cells in each waveguide was 25.

Band Diagram Calculations (MEEP)

The photonic band diagrams for the antislot unit cells were calculated with MEEP using the finite-difference time-domain (FDTD) method. To excite the TE band, a Gaussian beam with electric field polarized along the y-direction was used. For the LE band, a Gaussian beam with electric field polarized along the x-direction was applied. The hybridized bands were excited using a z-polarized magnetic field. The simulations utilized a 25 nm mesh size. The unit cell geometry consisted of a 220 nm-thick silicon layer modeled as suspended (substrate omitted to reduce computational cost), 400 nm lattice period, 150 nm radius, 700 nm unit cell height and 70 nm antislot bar width.

Transmission and Far-field Simulations (Ansys Lumerical FDTD)

The transmission spectra and far-field scattering patterns of the band-edge modes of the different antislot PhC waveguides were computed using 3D FDTD simulations with the same dimensions and number of unit cells as the fabricated samples. For each antislot rotation angle, both the transmission and the far-field projection profiles were obtained from a single simulation file with a simulation time of 100,000 fs. The mesh consisted of an automatic non-uniform grid in the x-y plane with an accuracy factor of 1, covering a 304 nm × 304 nm square region, and a 10 nm × 10 nm override mesh applied to the circular air hole region. A fundamental TE mode source was launched at the input side of each antislot PhC waveguide, and a frequency-domain power monitor was positioned on the output side in the y-z plane. To calculate the far-field properties, a near field frequency-domain field profile monitor captured the electric and magnetic field components approximately 165 nm above the antislot PhC waveguide. The recorded near-field data were then transformed into the far-field intensity distribution using the built-in far-field projection algorithm in Lumerical. The polarization-resolved far-field patterns were extracted by isolating the $E_x$ and $E_y$ components of the near field electric field and independently projecting them into the far field, yielding the polarization-decomposed far field intensity profiles.

Multipole Decomposition Simulation (COMSOL Multiphysics)

Each antislot PhC unit cell was treated as a meta-atom, and its optical response was determined using the multipole expansion method as detailed in Supplementary Information Note 4. Full-wave electromagnetic simulations were performed to obtain the forward and backward scattering power of each meta-atom. The numerical simulations were carried out using the finite element method (FEM) implemented in COMSOL Multiphysics. In particular, we used the wave optics module to solve Maxwell's equations in the frequency domain. For the chain of finite meta-atoms, we used a spherical air-filled domain with a radius of 4λ as the background medium. At the same time, perfectly matched layers of thickness 0.6λ were positioned outside of the background medium to act as absorbers and avoid back-scattering. A tetrahedral mesh was chosen to improve accuracy and facilitate numerical convergence. The meta-atoms were studied under plane wave illumination along the x-axis with an electric field pointing along the y-axis. After the calculations of the multipole moments, the scattered fields were then calculated in both forward and backward directions.

Transmission Measurements

Transmission measurements were performed using an edge-coupling configuration with TE-polarized light from a supercontinuum laser (NKT Photonics) covering the wavelength range of 800 to 2500 nm. The light source was filtered using an absorptive neutral density filter (NE10A-B, Thorlabs) and coupled from free space into a polarization-maintaining single-mode fiber (P1-1550PM-FC-2, Thorlabs) using a plano-convex lens (LA1951-AB-ML, Thorlabs). Customized tapered fibers (TPMJ-3U-1550-8/125-0.25-10-2.5-14-1, OZ Optics) were then used to couple the light into the antislot PhC waveguides. Polarization control was maintained by connecting the tapered fibers to single mode polarization-maintaining (PM) fibers through fiber-to-fiber U-benches (FBP-A-FC, Thorlabs). One side of the U-bench was connected with the tapered fiber, while the other side was connected with the single mode PM fiber (P1-1550PM-FC-2, Thorlabs). To interface with the spectrometer, a fiber adaptor (ADAFCSMA1, Thorlabs) was used to connect the PM fiber to an SMA fiber (QP1000-2-VIS-NIR, Ocean Optics) compatible with the fiber-coupled spectrometer (NIR Quest, Ocean Optics) operating over 900-1700 nm.

Far-field Scattering Measurements

The far-field scattered optical profile was recorded using an InGaAs camera (WiDy SWIR 640U-S, New Imaging Technologies) equipped with a 20x, long-working-distance objective (Mitutoyo Plan Apo NIR, NA=0.40, 20 mm working distance). A linear polarizer (LPNIR100, Thorlabs) mounted on a continuous rotation mount (CLR1, Thorlabs) was placed before the camera using a C-Mount to SM1 adapter (SM1A9, Thorlabs) to enable polarization resolved detection of the far-field scattered light. Measurements were conducted under consistent and controlled conditions: a fixed camera exposure time of 20 ms, an input laser power of -4 dBm, and minimized ambient illumination. For each device, far-field images were acquired at a wavelength corresponding to a band-edge mode and an off-band-edge mode identified from the transmission spectrum characterization. For polarization analysis, each band-edge mode image was captured at two orthogonal polarizations—0 degrees and 90 degrees relative to the waveguide axis (light propagation direction). The minimum step size of the rotation mount was 2°, which set the angular tolerance for polarization alignment in the measurements. To minimize the contributions of light scattering from the interface between the input waveguide and the first PhC unit cell, each band-edge mode image was subtracted from the corresponding off-band-edge mode image taken under the same polarization setting. Prior to acquiring device images, a cross-polarization calibration was performed to determine the absolute angular position of the rotation mount. Illumination from a known linearly polarized source was passed through a linear polarizer mounted on the rotation mount and imaged by the camera. For every 2 degree rotation angle, the total intensity was obtained by summing all pixel values in the recorded image under the same ambient light. The measured intensity follows a sinusoidal dependence on the analyzer angle, with the maximum intensity corresponding to alignment of the polarizer transmission axis with the incident polarization and the minimum intensity indicating the cross-polarized (orthogonal) orientation. The rotation mount readouts at these two positions define the absolute polarization reference angles. This calibration was used to analyze the far-field polarization properties of the antislot PhC waveguide.

**Data availability** The data that support the findings of this study are available from the corresponding authors upon reasonable request.

**Acknowledgement** This work was supported in part by the National Science Foundation (ECCS1407777, ECCS1809937, ECCS2240562). Antislot photonic crystals were fabricated at the Center for Nanophase Materials Sciences (CNMS), which is a US Department of Energy, Office of Science User Facility at Oak Ridge National Laboratory. SEM imaging was carried out in the Vanderbilt Institute of Nanoscale Science and Engineering (VINSE). T. Hong is thanked for helpful discussion and guidance on the MATLAB script used in the far-field imaging analysis. J. C. Ndukaife, G. Zhu, and I. Hong are acknowledged for providing access and technical assistance with the supercontinuum laser. J. D. Caldwell and G. Zhu are thanked for helpful discussion on coupled modes.

**Author contribution** S.H. and S.M.W. conceived the project. Y.Z. performed the band structure, field distribution, transmission, and far-field scattering simulations with assistance from S.H. and C.S.W. H.B.S. performed the multipole decomposition simulations. C.S.W. performed the coupled mode theory calculations. Y.Z. and C.S.W. fabricated the photonic crystals. Y.Z. performed the measurements. Y.Z., H.B.S., C.S.W., N.M.L., S.H., and S.M.W. analyzed the data and wrote the paper.

**Competing interests** The authors declare no competing interests

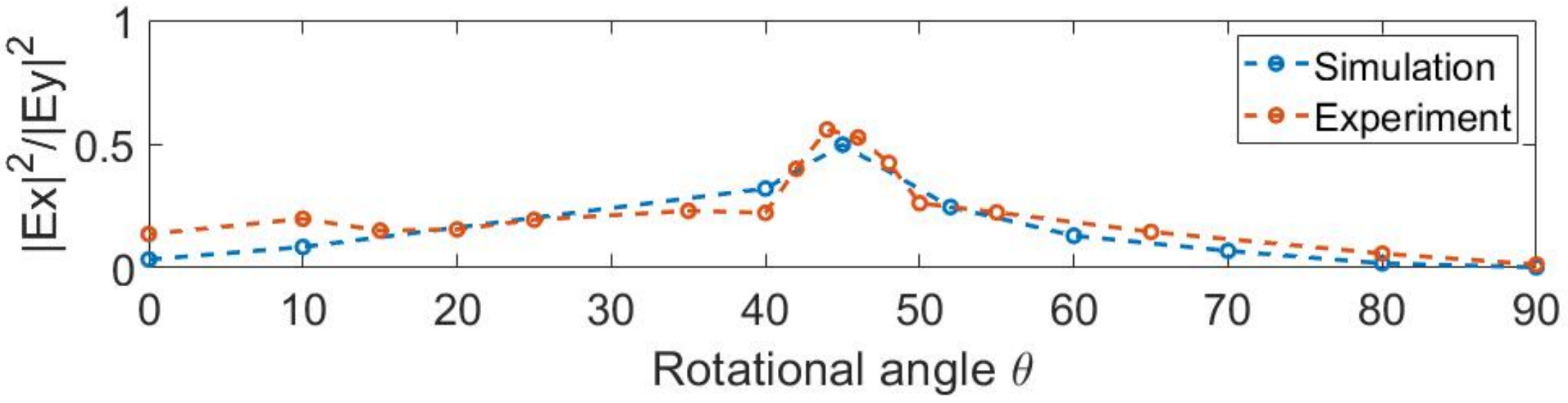

**Extended Data Fig. 1** | Ratio of the squared electric field in the $x$-direction, $|E_x|^2$ , to the squared electric field in the $y$-direction, $\left|E_y\right|^2$, integrated over the far-field profile at the fundamental resonance wavelength of antislot PhC waveguides with different antislot rotation angles, from far-field scattering simulations and experimental measurements. Results from simulation (blue dashed line, guide to the eye) and experiment (orange dashed line, guide to the eye) are in good agreement. For the R45 antislot PhC waveguide, the ratio reaches approximately 50%.

# Supplementary Information

## Supplementary Notes 1 to 4

**Supplementary Note 1 | Origin of longitudinal electric (LE) field distribution using Gauss's law**

To study the phase relationship between $E_x$ and $H_z$ ($x$ refers to the propagation direction along the waveguide, as shown in Supplementary Fig. S1), it is sufficient to examine the phase between $E_x$ and $E_y$, since net energy transport requires $E_y$ and $H_z$ to be in phase. Therefore, when $E_x$ is found to be in phase with $E_y$, it can be concluded that $E_x$ is also in phase with $H_z$ , allowing it to contribute to the energy flow along the waveguide.

By knowing the transverse component of the electric field, $\boldsymbol{E}$, $(E_y, E_z)$, we can express the longitudinal component $E_x$ from Gauss's law[1]:

$$\nabla \cdot (\varepsilon \boldsymbol{E}) = 0 \qquad (S1)$$

Eqn. S1 can be expanded to explicitly show the transverse and longitudinal components.

$$\nabla_t \cdot (\varepsilon \boldsymbol{E_t}) + \frac{\partial \varepsilon}{\partial x} E_x + \varepsilon \frac{\partial E_x}{\partial x} = 0 \qquad (S2)$$

In a homogeneous waveguide for which the permittivity is uniform in all directions, $\varepsilon(r) = \varepsilon(x, y, z)$ is a constant function such that $\frac{\partial \varepsilon}{\partial x} = \frac{\partial \varepsilon}{\partial y} = \frac{\partial \varepsilon}{\partial z} = 0$. Assuming $\boldsymbol{E}$ can be described as $\boldsymbol{E}(x, y, z) = \boldsymbol{E_t} e^{-i\beta x}$, Eqn. S2 can be simplified for a piecewise homogeneous waveguide to:

$$E_x = \frac{\nabla_t \cdot \boldsymbol{E_t}}{i\beta} \qquad (S3)$$

where $\beta = \frac{2\pi n_{eff}}{\lambda_0}$ and $n_{eff}$ is the effective index of the waveguide and $\lambda_0$ is the free space wavelength of light[2]. As indicated by the imaginary factor $i$ in Eqn. S3, $E_x$ is in quadrature phase (i.e., π/2 out of phase) with $\boldsymbol{E_t}$, as well as with the magnetic field ($H_z$), and therefore contributes no net energy flow through the waveguide.

However, in a 1D photonic crystal waveguide[3], the permittivity varies periodically along the propagation direction $x$. In this case, $\frac{\partial \varepsilon}{\partial x} \neq 0$. Therefore, the expression for $E_x$ from Eqn. S2 becomes:

$$E_x = A\left(\frac{\partial \varepsilon}{\partial x}(\nabla_t \cdot (\varepsilon \boldsymbol{E_t})) + i\varepsilon\beta(\nabla_t \cdot (\varepsilon \boldsymbol{E_t}))\right) \qquad (S4)$$

where

$$A = -\frac{1}{\left(\frac{\partial \varepsilon}{\partial x}\right)^2 + (\varepsilon\beta)^2} \qquad (S5)$$

Hence, in the case of waveguides with a permittivity that is not constant in the propagation direction, the longitudinal electric field $E_x$ has a real component (i.e., first term in Eqn. S4) and can therefore be partially in phase with $E_t$, as well as partially in-phase with the magnetic field, enabling it to contribute to net energy transport.

**Supplementary Note 2 | Origin of longitudinal electric (LE) field distribution using Ampère's law**

Here, we derive a general formula for the longitudinal electric field component ($E_x$) of light directly from Ampère-Maxwell's law:

$$\nabla \times \vec{H} = \vec{J} + \frac{\partial \vec{D}}{\partial t} \tag{S6}$$

Assuming the current density $\vec{J} = 0$ for dielectric waveguides, assuming harmonic time dependence $e^{i\omega t}$ for single frequency excitation, and rewriting the displacement field as $\vec{D} = \varepsilon \vec{E}$, we can express Eqn. S6 in the frequency domain:

$$\nabla \times \vec{H} = i\omega \vec{D} = i\omega\varepsilon \vec{E} \tag{S7}$$

Next, we expand the curl of $\vec{H}$ to show its components:

$$\left(\frac{\partial H_z}{\partial y} - \frac{\partial H_y}{\partial z}\right)\hat{x} + \left(\frac{\partial H_x}{\partial z} - \frac{\partial H_z}{\partial x}\right)\hat{y} + \left(\frac{\partial H_y}{\partial x} - \frac{\partial H_x}{\partial y}\right)\hat{z} = i\omega\varepsilon \vec{E} \tag{S8}$$

If we assume $H_z$ polarized light where $H_x = H_y \approx 0$, which is considered to be a TE polarization state in traditional waveguide platforms, Eqn. S8 simplifies to:

$$\left(\frac{\partial H_z}{\partial y}\right)\hat{x} - \left(\frac{\partial H_z}{\partial x}\right)\hat{y} = i\omega\varepsilon \vec{E} \tag{S9}$$

From Eqn. S9, we arrive at relationships that describe the well-known transverse $E_y$ component of light from the $H_z$ distribution, but also the longitudinal $E_x$ component of light from the $H_z$ distribution:

$$E_y = \left(-\frac{1}{i\omega\varepsilon}\right)\left(\frac{\partial H_z}{\partial x}\right) \tag{S10}$$

$$E_x = \left(\frac{1}{i\omega\varepsilon}\right)\left(\frac{\partial H_z}{\partial y}\right) \tag{S11}$$

Equation S11 implies that all electromagnetic waves that have a gradient of the $H_z$ magnetic field along the $y$ direction should have some $E_x$ contribution. Electromagnetic waves that are $H_z$ polarized but have no gradient along the $y$ direction, such as plane waves, do *not* have a longitudinal electric field component $E_x$. We note that in the phasor representation, all electromagnetic fields have a real and an imaginary component that together describe the amplitude and phase of the field. Therefore, the real and imaginary components of $H_z$ should always be considered when utilizing the equations given above.

Finally, we demonstrate the agreement of Eqn. S11 with FDTD simulation results from Ansys Lumerical in Fig. S1. Here, we consider antislot PhC waveguides with different antislot rotations angles. These results are obtained from unit cell simulations employing Bloch boundary conditions. Each unit cell is excited by an electric dipole oriented along the $y$ direction. Electric and magnetic field distributions in the $(x, y)$ plane are taken from the center of the silicon waveguide along the $z$ direction. The $H_z$ and $E_x$ field components are taken directly from simulation, while the $\partial H_z/\partial y$ distribution is calculated from the $H_z$ field distribution. The good agreement between the calculated $\partial H_z/\partial y$ and simulated $E_x$ distributions is consistent with Eqn. S11 and highlights that the $H_z$ magnetic field component is a dominant factor in determining the strength of the longitudinal electric field.

**Supplementary Note 3 | Coupled mode theory describes band hybridization and anticrossing**

Coupled mode theory provides a strong framework to describe the TE-LE hybridization quantitatively from a band structure perspective[4,5]. Equation S12 shows the eigenvalue problem used to quantify the hybridization between the pure TE and pure LE photonic modes:

$$\begin{bmatrix} f_{TE} & g_{TE-LE} \\ g_{TE-LE} & f_{LE} \end{bmatrix} \begin{bmatrix} \alpha_{TE} \\ \alpha_{LE} \end{bmatrix} = f_{hybrid} \begin{bmatrix} \alpha_{TE} \\ \alpha_{LE} \end{bmatrix} \tag{S12}$$

Here, $f_{TE}$ and $f_{LE}$ represent the frequencies of the pure TE and pure LE bands, $f_{hybrid}$ represents the frequency of the hybrid band, $g_{TE-LE}$ (or $g$) represents the coupling strength between the pure TE and LE bands, and $\alpha_{TE}$ and $\alpha_{LE}$ are typically referred to as the mixing coefficients. By convention, $|\alpha_{TE}|^2 + |\alpha_{LE}|^2 = 1$. Each variable is a function of $k_x$, the wave vector, and as such, the eigenvalue problem is solved numerically for each $k_x$. Finally, this eigenvalue problem will have two solutions, for $f_{hybrid} = f_{UH}$ and $f_{hybrid} = f_{LH}$, where $UH$ represents the upper hybrid band and $LH$ represents the lower hybrid band.

Unlike many systems, we are unable to decouple our system to obtain the pure TE and pure LE bands ($f_{TE}$ and $f_{LE}$) when the antislot is rotated away from its R0 and R90 positions. R0 and R90 are uncoupled systems, but their pure bands differ (i.e., $f_{TE(R0)} \neq f_{TE(R90)}$ and $f_{LE(R0)} \neq f_{TE(R90)}$). To address this issue, we approximate the pure TE and pure LE bands for intermediate angles by linearly interpolating fits to the R0 uncoupled bands and to the R90 uncoupled bands. We found that a fifth-order polynomial is adequate to model this behavior (Eqn. S13). This approach is empirically determined to be sufficient as the simulated crossings or degeneracy points are at the midpoint of the novel bandgap. The fits for the pure TE and LE bands are displayed in Table S1 and S2, respectively.

$$f_{pure} = p_1(k_x)^5 + p_2(k_x)^4 + p_3(k_x)^3 + p_4(k_x)^2 + p_5(k_x) + p_6 \tag{S13}$$

Now that we have a suitable approximation for the pure TE and LE bands, we can begin solving the eigenvalue problem for $g$, the coupling strength coefficient. At each $k_x$, we can solve the determinant as follows:

$$\det|A - \lambda I| = 0 = \det\left|\begin{bmatrix} f_{TE} & g \\ g & f_{LE} \end{bmatrix} - \begin{bmatrix} f_{hybrid} & 0 \\ 0 & f_{hybrid} \end{bmatrix}\right|$$

$$= \det\left|\begin{bmatrix} f_{TE} - f_{hybrid} & g \\ g & f_{LE} - f_{hybrid} \end{bmatrix}\right|$$

$$0 = (f_{TE} - f_{hybrid})(f_{LE} - f_{hybrid}) - g^2$$

$$g = \sqrt{(f_{TE} - f_{hybrid})(f_{LE} - f_{hybrid})}$$

$$g = \sqrt{f_{TE}f_{LE} - f_{TE}f_{hybrid} - f_{LE}f_{hybrid} + f_{hybrid}^2} \tag{S14}$$

Here, we can use the computed frequencies for the hybrid bands from our band structure simulations. Due to the approximations in determining the pure TE and pure LE bands for intermediate angles, the $g$ coefficient determined for the upper and lower hybrid bands differs slightly. We account for this by averaging $g_{UH}$ (upper hybrid band) with $g_{LH}$ (lower hybrid band) to obtain $g(k_x)$ for a given rotation angle. Physically, $g$ represents the deviation from the fitted pure bands and therefore has units of frequency. $g$ is most intuitively understood at the anti-crossing point; there, $g$ is equal to half the bandgap size. Figure S2 shows the computed $g$ at the crossing or anti-crossing point for all antislot angles.

By solving this eigenvalue problem using software such as MATLAB, we also reproduce our eigenvalue – the frequency of the hybrid bands. Figure S3 plots the band structure simulation data, along with the assumed pure TE and pure LE bands, and the recalculated hybrid bands that utilize our computed $g$ coefficient. The close match between the hybrid band structure simulations with the recomputed hybrid band fit demonstrates both that the approximated pure TE and pure LE bands are adequate, but also that the $g$ coefficient alone is sufficient to describe the coupling behavior between the two modes.

In the same step, the eigenvectors $\alpha_{TE}$ and $\alpha_{LE}$ (mixing coefficients) were determined. This mixing coefficient represents the fraction of the pure TE or LE band one of the hybrid bands is at each point in $k_x$ space. Because only two modes are coupling with one another, these coefficients are reciprocal (i.e., if the upper hybrid band is 30% TE and 70% LE, the lower hybrid band will be 70% TE and 30% LE). Knowledge of the mixing fraction of the hybrid modes does not indicate more or less longitudinal electric field component – rather, it gives an indication of which mode profile (pure TE or pure LE) the hybrid mode will resemble more at a specific $k_x$ vector. In addition, the slope of the mixing coefficients indicates how the two modes are coupling – if it is a quick exchange of energy (low coupling) or if the coupled mode spends more time in $k_x$ space as a combination of the two modes (higher degree of coupling). The mixing coefficients for the upper hybrid band are shown in Figure S4 (the mixing coefficients are inverted for the lower hybrid band).

**Supplementary Note 4 | Multipole Decomposition**

The optical response of nonmagnetic meta-atoms is characterized by the electric current density $\boldsymbol{J}_{in}$ and polarization $\boldsymbol{P}_{in}$ induced by external electromagnetic fields. In the case of monochromatic fields, these values are connected by $\boldsymbol{J}_{in} = -i\omega \boldsymbol{P}_{in}$, wherein $\omega$ represents the angular frequency. Upon the interaction of light with the meta-atom, the induced polarization is related to the field distributions within the particle via $\boldsymbol{P}_{in} = \epsilon_0(\epsilon_p - \epsilon_d)\boldsymbol{E}_p$, where $\epsilon_0$, $\epsilon_p$ and $\epsilon_d$ are the free space, meta-atom, and surrounding medium dielectric constants, respectively, and $\boldsymbol{E}_p$ is the total electric field inside the meta-atom[6–8]. Therefore, the internal field distribution can directly change the induced polarization, which in turn alters the excited moments within the meta-atom. In the far-field region, the scattered field can be expressed as

$$\boldsymbol{E_n}(\boldsymbol{r}) = i\omega\mu_0 \frac{\exp(ik_d r)}{4\pi r}\left(\bar{\bar{I}} - \boldsymbol{nn}\right)\int_{V_s} \boldsymbol{J}_{in}(\boldsymbol{r}')\exp\left(-ik_d(\boldsymbol{n}\cdot\boldsymbol{r}')\right)d\boldsymbol{r}' \quad \text{(S15)}$$

where $\boldsymbol{n} = \boldsymbol{r}/r$ is the unit vector directed from the particle's center towards an observation point, $k_d$ is the wavenumber in the surrounding medium, and $\hat{\hat{I}}$ is a $3 \times 3$ unitary matrix. The multipole decomposition of scattered waves can be obtained based on three approaches: (a) Taylor expansion of $\exp(-ik_d r)$ around a point with the radius vector $\boldsymbol{r_0}$ located in the meta-atom volume; (b) writing the induced current density within the meta-atom in terms of a Kronecker delta function as $\boldsymbol{J}_{in}(\boldsymbol{r}') = \int_{V_s}\boldsymbol{J}_{in}(\boldsymbol{r})\delta(\boldsymbol{r}' - \boldsymbol{r})d\boldsymbol{r}'$, and then performing a Taylor expansion of the $\delta(\boldsymbol{r}' - \boldsymbol{r})$ term in the vicinity of point $\boldsymbol{r_0}$ and (c) spherical harmonic expansion of the $\exp(-ik_d r)$ term directly. As was shown in references 6-8, the first two methods ((a) and (b)) lead to the same results and provide multipoles in the long wavelength approximation (LWA) regime, while the third approach yields the multipole decomposition of the scattered field by finite-size scatterers with arbitrary dimensions, providing the exact multipole moments. To obtain the exact moments, the plane wave representation of $\exp\left(-ik_d(\boldsymbol{n}\cdot\boldsymbol{r}')\right)$ in the spherical harmonics leads to

$$\exp\left(-ik_d(\boldsymbol{n}\cdot\boldsymbol{r}')\right) = 4\pi\sum_{l=0}^{\infty}\sum_{m=-l}^{l}(-i)^l j_l(k_d r')Y_{lm}^*(\theta,\varphi)Y_{lm}(\theta,\varphi) \quad \text{(S16)}$$

wherein $Y_{lm}(\theta,\varphi)$ represents the spherical harmonics, $j_l(k_d r')$ is the spherical Bessel function of order $l$, and the asterisk (*) denotes the complex conjugation. Substituting Eq. (S15) into Eq. (S16), the scattered electric field can be obtained as

$$\boldsymbol{E_n}(\boldsymbol{r}) = i\omega\mu_0 \frac{\exp(ik_d r)}{4\pi r}\left(\bar{\bar{I}} - \boldsymbol{nn}\right)\sum_{l=0}^{\infty}(-i)^l(2l+1)\int_{V_s}\boldsymbol{J}_{in}(\boldsymbol{r}')P_l(\cos(\gamma))d\boldsymbol{r}' \quad \text{(S17)}$$

where $P_l(\cos(\gamma))$ is the $l$-order Legendre polynomial, obtained from the application of the addition theorem for spherical harmonics. The induced multipole moments (up to electric quadrupole term) within the meta-atom can then be directly derived based on different combinations of $l$ as follows

$$\boldsymbol{D} = \int j_0(k_0 r')\boldsymbol{P} d\boldsymbol{r}' + \frac{k_d^2}{10}\int \frac{15 j_2(k_d r')}{(k_d r')^2}\left[(\boldsymbol{r}' \cdot \boldsymbol{P})\boldsymbol{r}' - \frac{1}{3}r'^2\,\boldsymbol{P}\right] d\boldsymbol{r}',$$

$$\boldsymbol{m} = -\frac{3}{2}i\omega \int \frac{j_1(k_d r')}{k_d r'}[\boldsymbol{r}' \times \boldsymbol{P}]\, d\boldsymbol{r}',$$

$$\widehat{M} = -5i\omega \int \frac{j_2(k_d r')}{(k_d r')^2}([\boldsymbol{r}' \times \boldsymbol{P}] \otimes \boldsymbol{r}' + \boldsymbol{r}' \otimes [\boldsymbol{r}' \times \boldsymbol{P}])d\boldsymbol{r}' \tag{S18}$$

$$\hat{Q} = \int \frac{3 j_1(k_d r')}{k_d r'}\left[3(\boldsymbol{r}' \otimes \boldsymbol{P} + \boldsymbol{P} \otimes \boldsymbol{r}') - 2(r' \cdot \boldsymbol{P})\bar{\bar{I}}\right]d\boldsymbol{r}'$$
$$+ 6k_d^2 \int \frac{j_3(k_d r')}{(k_d r')^3}\left[5\,\boldsymbol{r}' \otimes \boldsymbol{r}'(\boldsymbol{r}' \cdot \boldsymbol{P}) - r'^2(\boldsymbol{P} \otimes \boldsymbol{r}' + \boldsymbol{r}' \otimes \boldsymbol{P}) - (\boldsymbol{P} \cdot \boldsymbol{r}')r'^2\bar{\bar{I}}\right]d\boldsymbol{r}'$$

where $\boldsymbol{D}$ corresponds to the exact total electric dipole (TED), $\boldsymbol{m}$ is the exact magnetic dipole (MD) moment, and $\hat{Q}$, and $\widehat{M}$ represent the electric quadrupole (EQ), and magnetic quadrupole tensors (MQ). The operators of $\cdot$ , $\times$, and $\otimes$ represent the scalar, vector, and tensor products, respectively.

# Supplementary Tables 1 to 3

**Supplementary Table 1 |** Fifth order polynomial fits for the pure TE band. Only R0 and R90 were decoupled and could be directly fitted ($R^2$ values shown for direct fits). Intermediate angle fits were determined by linearly interpolating the coefficients between the R0 and R90 fits.

| Antislot PhC Waveguide | $p_1$ | $p_2$ | $p_3$ | $p_4$ | $p_5$ | $p_6$ | $R^2$ |
|---|---|---|---|---|---|---|---|
| **R0** | -118.59 | 140.70 | -17.90 | -37.80 | 17.86 | -2.03 | 0.9966 |
| **R15** | 139.71 | -372.56 | 386.07 | -195.04 | 48.06 | -4.32 | -- |
| **R30** | 398.00 | -885.81 | 790.04 | -352.28 | 78.26 | -6.60 | -- |
| **R45** | 656.29 | -1399.07 | 1194.01 | -509.51 | 108.45 | -8.89 | -- |
| **R60** | 914.58 | -1912.32 | 1597.98 | -666.75 | 138.65 | -11.17 | -- |
| **R75** | 1172.87 | -2425.58 | 2001.95 | -823.99 | 168.85 | -13.46 | -- |
| **R90** | 1431.17 | -2938.83 | 2405.92 | -981.23 | 199.05 | -15.75 | 0.9977 |

**Supplementary Table 2 |** Fifth order polynomial fits for the pure LE band. Only R0 and R90 were decoupled and could be directly fitted ($R^2$ values shown for direct fits). Intermediate angle fits were determined by linearly interpolating the coefficients between the R0 and R90 fits.

| Antislot PhC Waveguide | $p_1$ | $p_2$ | $p_3$ | $p_4$ | $p_5$ | $p_6$ | $R^2$ |
|---|---|---|---|---|---|---|---|
| **R0** | 25.60 | -69.52 | 70.14 | -33.79 | 8.09 | -0.51 | 1.0000 |
| **R15** | 40.29 | -99.66 | 94.77 | -43.86 | 10.15 | -0.68 | -- |
| **R30** | 54.98 | -129.78 | 119.39 | -53.93 | 12.21 | -0.84 | -- |
| **R45** | 69.67 | -159.94 | 144.01 | -64.00 | 14.26 | -1.01 | -- |
| **R60** | 84.36 | -190.08 | 168.63 | -74.07 | 16.32 | -1.17 | -- |
| **R75** | 99.04 | -220.22 | 193.25 | -84.14 | 18.38 | -1.34 | -- |
| **R90** | 113.73 | -250.36 | 217.87 | -94.21 | 20.44 | -1.50 | 1.0000 |

**Supplementary Table 3** | Comparison of properties related to longitudinal electric fields in different free space and guided-wave systems.

| | System Type | $\frac{\partial \varepsilon}{\partial x} \neq 0$ | Significant LE field component | Hybrid LE-TE mode | References |
|---|---|---|---|---|---|
| **Free space** | Laser beam | **No** | **No** | **-** | - |
| | Focused laser beam | **Yes** | **Yes** | **-** | e.g., [9] |
| **Guided wave** | Ridge waveguide | **No** | **No** | **No** | - |
| | Nanowire waveguide | **No** | **Yes** | **No** | e.g., [2] |
| | Traditional PhC waveguide | **Yes** | **No** | **No** | - |
| | Rotated antislot PhC waveguide | **Yes** | **Yes** | **Yes** | This work |

Note: $x$ represents propagation direction along the waveguide

# Supplementary Figures 1 to 4

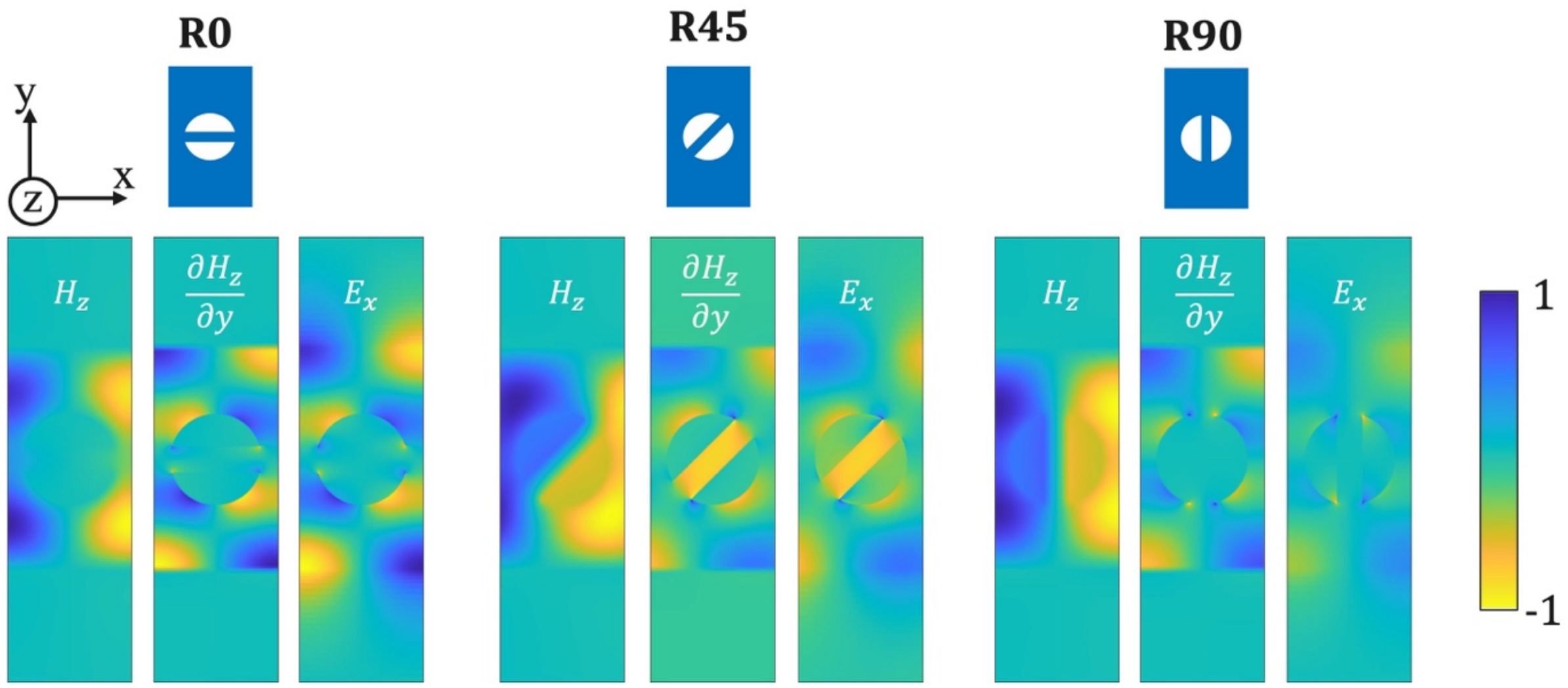

**Supplementary Fig. S1 |** Comparison of different antislot PhC waveguide unit cells with their simulated $H_z$ distribution, calculated $\partial H_z/\partial y$, and simulated $E_x$ distribution. The $H_z$ and $E_x$ distributions are direct simulation results. The $\delta H_z/\delta y$ distribution is computed using only the $H_z$ simulation data and is proportional to the $E_x$ simulation results, as seen by the matching spatial field distributions. This good agreement demonstrates that $H_z$ is a key contributing factor to the $E_x$ distribution.

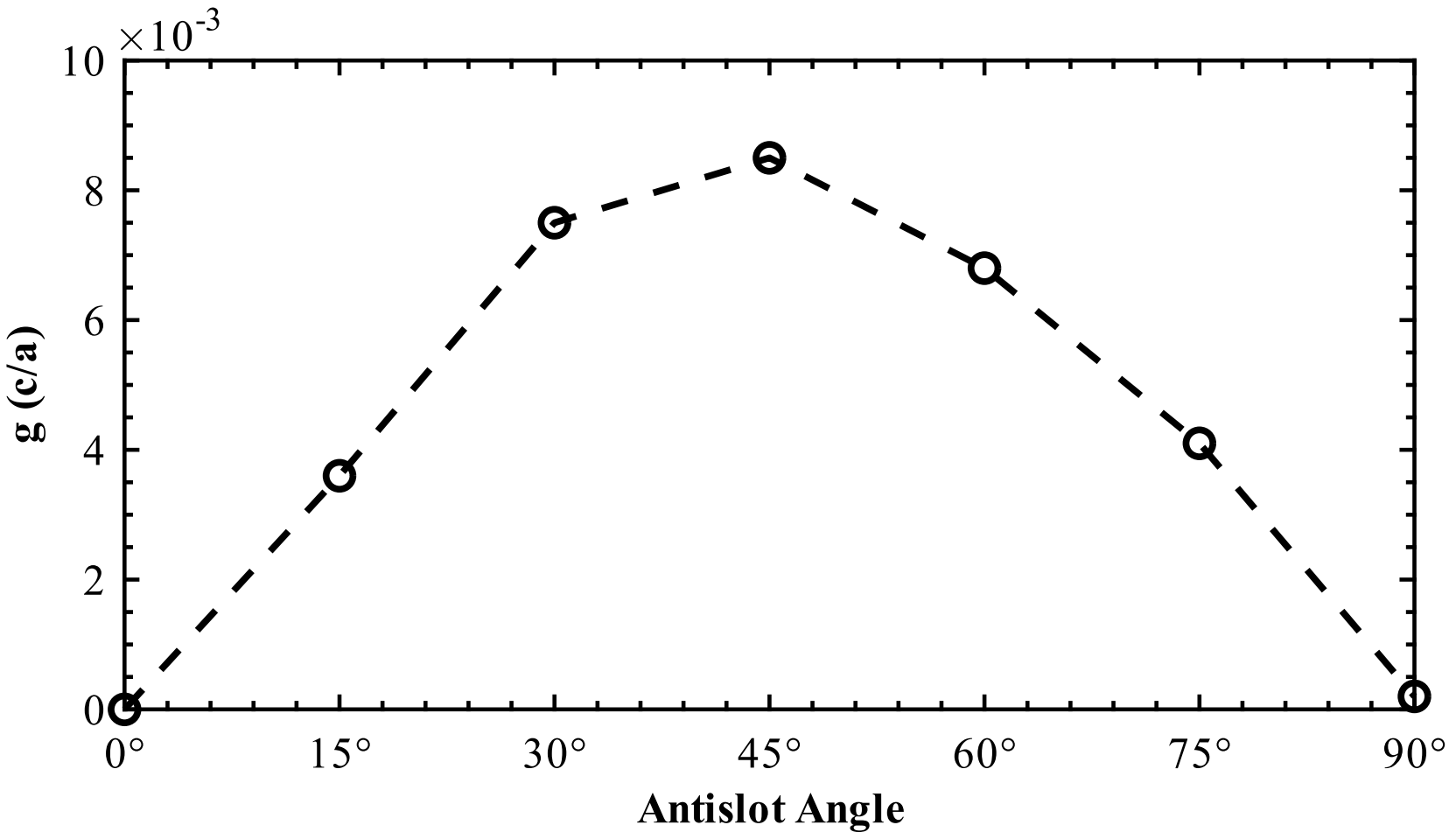

**Supplementary Fig. S2 |** Coupling strength coefficient as a function of antislot angle at each respective anti-crossing point ($g(k_x =$ anti-crossing)). At the anti-crossing point, $g$ physically represents half the size of the bandgap.

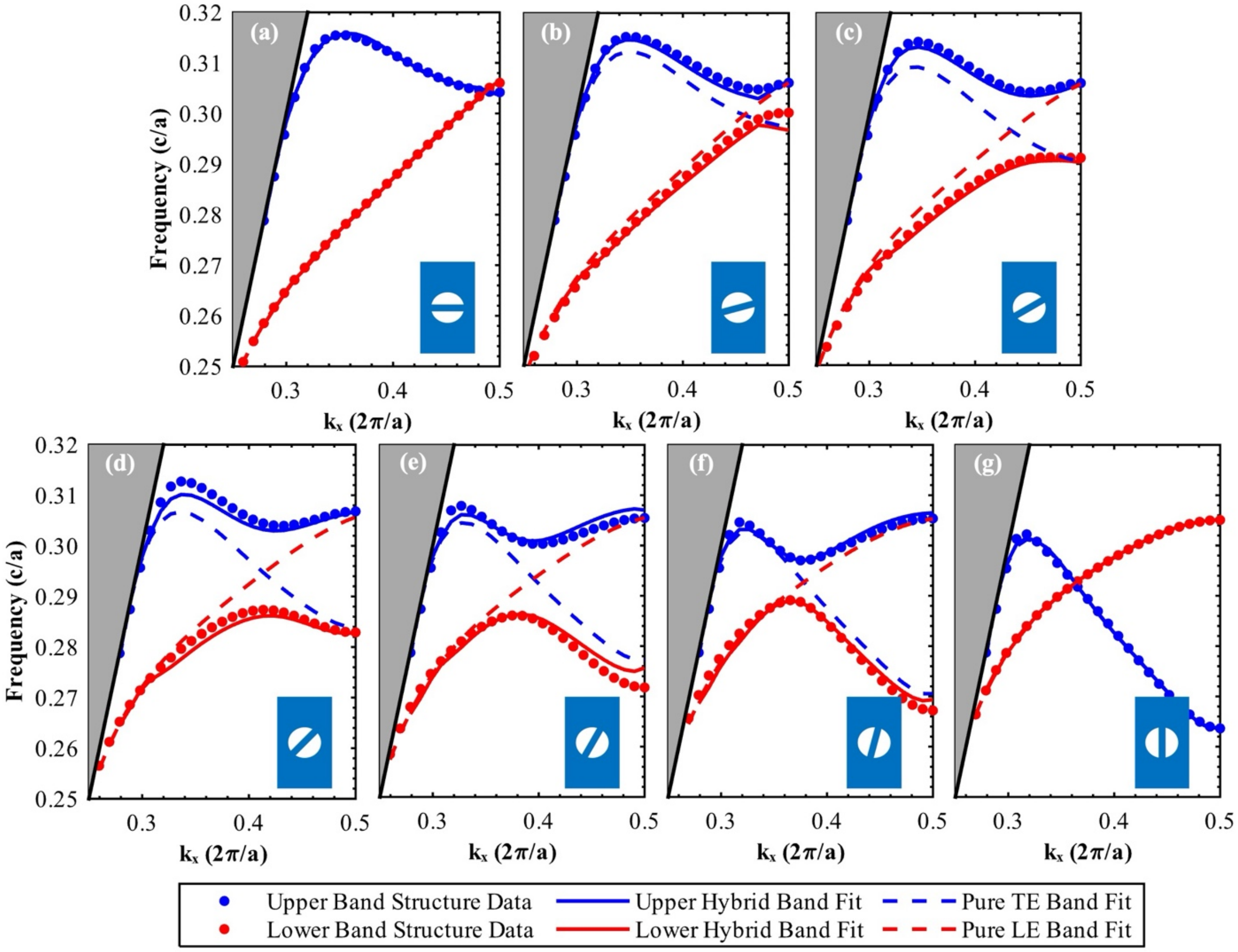

**Supplementary Fig. S3 |** Band structure data and fits for (a) R0, (b) R15, (c) R30, (d) R45, (e) R60, (f) R75, and (g) R90. The circle markers represent data that was determined via band structure simulations (i.e., MPB). The dashed lines represent the assumed uncoupled pure TE and pure LE bands that are approximated using the fits from Tables S1 and S2. The solid lines represent the coupled mode theory fit to describe the hybridization – this fit utilizes only the approximated pure TE and LE modes and the average fitted $g$ coefficient. Close matching between this fit and the simulated band structure data indicates that this model adequately describes the band hybridization.

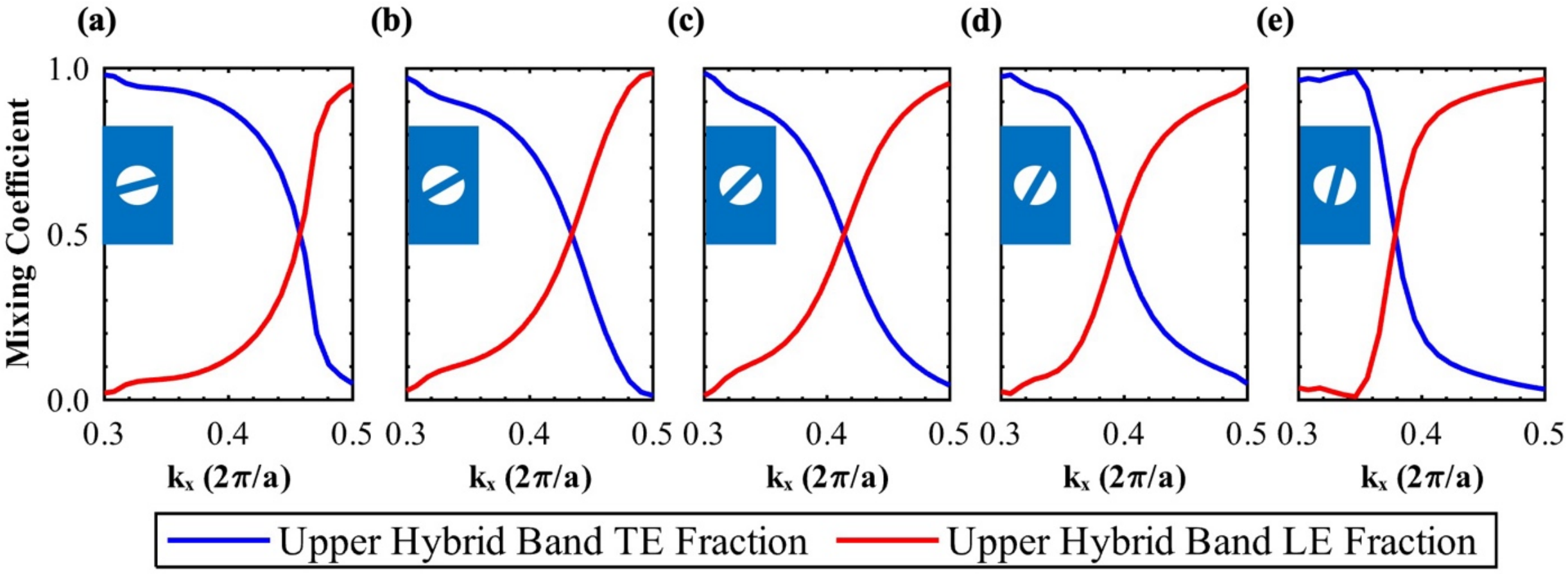

**Supplementary Fig. S4 |** Mixing coefficients for the upper hybrid band for (a) R15, (b) R30, (c) R45, (d) R60, and (e) R75. R0 and R90 are not shown because an upper hybrid band and lower hybrid band are not well-defined for an uncoupled system. The closer to R45, the longer in $k_x$ the hybrid mode stays significantly mixed.

**Supplementary information references**